\documentclass[final,1p,times]{elsarticle} 
\usepackage{graphicx} 
\usepackage{amssymb} 
\usepackage{amsthm} 

\newcommand{\hla}{$\mathrm{^{3}_{\Lambda}H}$}

\newcommand{\sNN}{$\sqrt{s_{_\mathrm{NN}}}$ }
\newcommand{\he}{$\mathrm{^{3}He}$}

\journal{Nuclear Physics A} 

\begin{document} 

\begin{frontmatter} 


\title{Observation of hypertritons in Au+Au collisions at \sNN = 200~GeV}

\author{J H Chen$^1$$^,$$^2$ for the STAR
Collaboration} \ead{jhchen@rcf.rhic.bnl.gov}
\address{$^1$Physics department, Kent State University, Kent, OH, 44242, USA}
\address{$^2$Division of nuclear physics, Shanghai Institute of Applied Physics, CAS, Shanghai, 201800, China}

\begin{abstract} 
We report preliminary results of \hla~observation in
heavy-ion collisions at RHIC. We have identified 157$\pm$30
candidates in the current sample containing
$\sim$$10^8$ Au+Au events at \sNN = 200~GeV. The production rate
of \hla~is close to that of \he. No extra penalty factor is
observed for \hla, in contrast to results observed at the
AGS.
\end{abstract} 

\end{frontmatter} 



\section{Introduction}\label{1}
Hypernuclear physics opens a unique opportunity for the study of
the hyperon-nucleon (YN) and the hyperon-hyperon (YY) interaction.
A hypernucleus has a non-zero strangeness quantum number, and
thus for nuclear spectroscopy, it provides one more degree of
freedom than a normal nucleus containing only protons and neutrons.
Information on the strangeness sector of the hadronic
equation of state is crucial for understanding the structure of a
neutron star. Depending on the strength of the YN interaction, a
neutron star might be a hyperon star, or strange quark matter, or
might have a kaon condensate at its core~\cite{neuteronstar}. The
lightest and simplest hypernucleus is the hypertriton (\hla),
consisting of a Lambda, a proton and a neutron. The hypernucleus
was first observed in 1952~\cite{hyperT1953}, while no
anti-hypernucleus has been found yet.

The Relativistic Heavy-Ion Collider produces a system that
consists of a large number of particles with high phase-space
density, and with almost equal numbers of quarks and anti-quarks.
This environment is uniquely suited for production of exotic
nuclei, including hypernuclei, anti-nuclei, and anti-hypernuclei.

\section{Analysis and Results}\label{2}

In this paper, we report STAR preliminary results on
\hla~measurement in Au+Au collisions at \sNN=200~GeV. The signal
candidates were reconstructed in the central TPC~\cite{TPC2003}
and were identified via the secondary vertex of \hla~$\rightarrow
 \mathrm{^{3}He + \pi^{-}}$. Approximately $\mathrm{2.3
\times 10^7}$ minimum-bias (MB) trigger events and $\mathrm{2.2
\times 10^{7}}$ central trigger events from Au + Au data
collected in the year 2004 run, and $\mathrm{6.8 \times 10^7}$ MB
events in the year 2007 Au+Au run have been used in this analysis.
Events were required to have a primary vertex position within 30
cm from the center of the TPC along the beam direction. Charged
tracks were reconstructed in the STAR TPC for pseudorapidity
$|\eta|<1.8$ with full azimuthal acceptance ($\mathrm{0 \leq \phi
\leq 2\pi}$).

\begin{figure}[htb]
\includegraphics[scale=0.72,bb=20 -35 100 280]{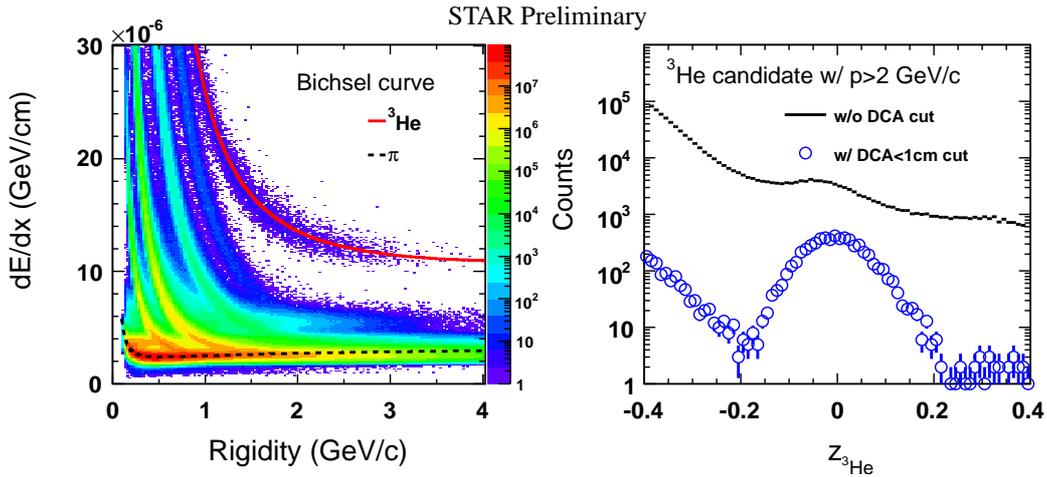}
\put(105,195){STAR Preliminary}
\vspace{-0.55cm}\caption{\label{fig1}The left panel shows $dE/dx$
versus rigidity for charge particles reconstructed in the STAR
TPC. Different bands represent different particle types. The
expected curves for \he~and $\pi$ are also plotted. The right
panel shows the $z_{\rm ^{3}He}$ distribution for \he~candidates
at $p > 2$ GeV$/c$. The black line represents the data without a
DCA cut, and the open circles show the result with
$\mathrm{DCA}<1$ cm.}
\end{figure}

Particle identification is achieved by correlating the ionization
energy loss $(dE/dx)$ of charged particles in the TPC gas with
their measured magnetic rigidity (Fig~\ref{fig1}, left panel).
Since the $\langle dE/dx \rangle$ distribution for a fixed
particle type is not Gaussian, a new variable $z_{i}$
($i=\pi,K,p,d,t,\, {\rm ^3He}$, ...) is useful in order to
facilitate deconvolution into Gaussians for each particle
species~\cite{PID2009}. The $z_{i}$ is defined as
$z_{i}=\ln(\langle dE/dx \rangle / \langle dE/dx
\rangle^{B}_{i})$, where $\langle dE/dx \rangle^{B}_{i}$ is the
expected value of $\langle dE/dx \rangle$ for the given particle
type~\cite{PDG2008}.  We are only able to identify \he~with larger
momenta (i.e., $p>2$~GeV$/c$). Furthermore, the beam pipe causes
contamination of the \he~sample. Since most beam pipe knock-out
\he~are far away from the collision vertex, a cut on distance of
closest approach to the collision vertex ($\mathrm{DCA}<1$ cm)
reduces the background associated with the beam pipe.
Fig~\ref{fig1} (right panel) shows the \he~candidate distributions
before and after the DCA cut. The condition
$\mathrm{|z_{^{3}He}|<0.2}$ with the 1-cm DCA cut are used in this
analysis. The daughter pion from the \hla~decay usually has a
relatively low momentum, and it can be cleanly identified with
similar $\langle dE/dx \rangle$ selection in our experiment
(Fig~\ref{fig1}, left panel)~\cite{PID2009}.

With both daughter candidates identified, we can reconstruct the
signal through its weak decay topology. A set of topological cuts
has been applied here in order to reduce the combinatorial
background. The cuts include: DCA between \he~and pion tracks
($<1$ cm), DCA of the \hla~candidate to the primary vertex ($<1$
cm), decay length of the \hla~candidate vertex from the collision
vertex ($>2.4$ cm), and the DCA of the pion daughter to the
primary vertex ($>0.8$ cm). The cuts were optimized based on a
full detector response simulation~\cite{PID2009}. The secondary
vertex finding technique used in our experiment is essential for
this rare particle search because it can reduce the background
dramatically.

\begin{figure}[htb]
\includegraphics[scale=0.54,bb=-80 -35 240 460]{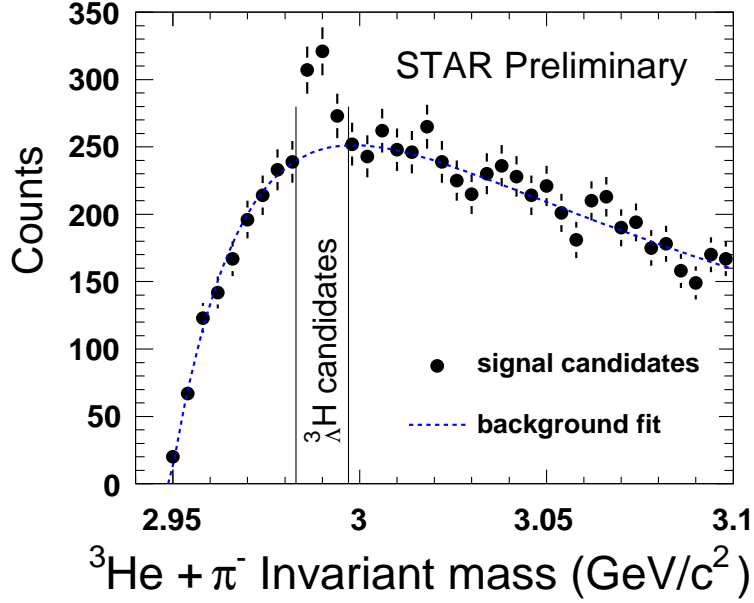}
\vspace{-0.55cm}\caption{\label{fig2}STAR preliminary invariant
mass distribution for the daughters $\mathrm{^{3}He + \pi^{-}}$ in
Au+Au collisions at \sNN=200~GeV. The solid circles represent the
signal candidate distribution, and the dashed line is the
corresponding fitted background distribution.}
\end{figure}

Fig.~\ref{fig2} shows the invariant mass distribution of
$\mathrm{^{3}He + \pi^{-}}$ with all cuts applied. The solid
circles represent the signal candidate distribution while the
dashed curve is the corresponding background distribution. The
signal candidate invariant mass was calculated based on the
momenta of the daughter candidates at the decay vertex:
$M^{2}=E^{2}-P^{2},~~ E=E_{\rm ^{3}He}+E_{\pi},~~
\overrightarrow{P}=\overrightarrow{P}_{\rm
^{3}He}+\overrightarrow{P}_{\pi}$. The background distribution is
produced by the event rotation method. Instead of directly
subtracting the background from the data, we use a fitting
function. Here we use a double exponential function: $f(x)\propto
\exp(-\frac{x}{p_{1}}) - \exp(-\frac{x}{p_{2}})$ to fit the
background distribution, where $p_1, p_2$ are fitting parameters.
The signal is then counted after subtraction of the fitted rotated
background. In total, 157$\pm$30 \hla~candidates were found in the
current data sample. A very similar analysis can be used to search
for anti-hypertritons decaying to $\mathrm{^{3}\bar{He} +
\pi^{+}}$. Results of this search are reported elsewhere.

\begin{figure}[htb]
\includegraphics[scale=0.36,bb=-220 -35 200 760]{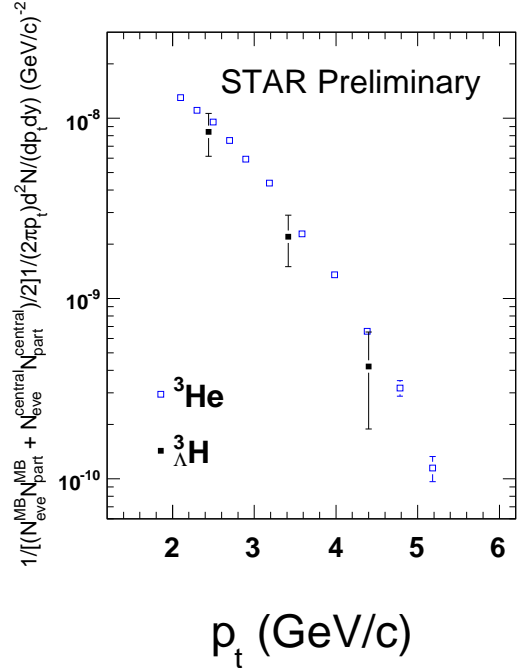}
\vspace{-0.55cm}\caption{\label{fig3}STAR preliminary results on
\hla~and \he~invariant yield vs. $p_t$. The results
have been corrected for detector acceptance and tracking
efficiency. The error bars represent statistical uncertainties
only.}
\end{figure}

Fig.~\ref{fig3} shows the preliminary \hla~$p_t$ spectra together
with the \he~results in the same data set. The results are
corrected for detector acceptance and tracking efficiency. Since
we combine the MB and central trigger data for this analysis, we
here normalize the invariant yield per event per participant pair:
$\mathrm{(N_{eve}^{MB} \times N_{part}^{MB} + N_{eve}^{central}
\times N_{part}^{central})/2}$ rather than the conventional number
of total events: $\mathrm{N_{eve}}$. We then compare the
\hla~production rate with \he~in our data. It is quite surprising
to observe that their production ratio is close to unity. This is
different from the value obtained at the AGS, where an extra
penalty factor has been observed when introducing the strangeness
degree of freedom in particle production~\cite{AGS2004}. It seems
that our results cannot be solely due to the finite size effect as
implied by AGS data~\cite{AGS2004}. In contrast, our measurements
indicate that at RHIC, the strangeness phase space population is
similar to that of light quarks.

\section{Summary}\label{3}
In summary, we present preliminary
\hla~measurements in Au+Au collisions at \sNN=200~GeV. We have
collected $157\pm30$ signal candidates in the current
statistics. Our measurement of the \hla/\he~ratio is consistent with the
enhancement of strangeness production at RHIC.



\bibliography{hypertriton-QM09}

\begin{thebibliography}{6}
\providecommand{\natexlab}[1]{#1}
\providecommand{\url}[1]{\texttt{#1}}
\providecommand{\urlprefix}{URL }
\expandafter\ifx\csname urlstyle\endcsname\relax
  \providecommand{\doi}[1]{doi:\discretionary{}{}{}#1}\else
  \providecommand{\doi}[1]{doi:\discretionary{}{}{}\begingroup
  \urlstyle{rm}\url{#1}\endgroup}\fi
\providecommand{\bibinfo}[2]{#2}

\bibitem[{Lattimer and Prakash(2004)}]{neuteronstar}
\bibinfo{author}{J.~M. Lattimer}, \bibinfo{author}{M.~Prakash},
  \bibinfo{journal}{Science} \bibinfo{volume}{304} (\bibinfo{year}{2004})
  \bibinfo{pages}{536}.

\bibitem[{Danysz and Pniewski(1953)}]{hyperT1953}
\bibinfo{author}{M.~Danysz}, \bibinfo{author}{J.~Pniewski},
  \bibinfo{journal}{Phil. Mag.} \bibinfo{volume}{44} (\bibinfo{year}{1953})
  \bibinfo{pages}{348}.

\bibitem[{Anderson et~al.(2003)}]{TPC2003}
\bibinfo{author}{M.~Anderson}, et~al., \bibinfo{journal}{Nucl. Instrum. Methods
  A} \bibinfo{volume}{499} (\bibinfo{year}{2003}) \bibinfo{pages}{659}.

\bibitem[{Abelev et~al.(2009)}]{PID2009}
\bibinfo{author}{B.~I. Abelev}, et~al., \bibinfo{journal}{Phys. Rev. C}
  \bibinfo{volume}{79} (\bibinfo{year}{2009}) \bibinfo{pages}{034909}.

\bibitem[{Amsler et~al.(2008)}]{PDG2008}
\bibinfo{author}{C.~Amsler}, et~al., \bibinfo{journal}{Phys. Lett.}
  \bibinfo{volume}{B667} (\bibinfo{year}{2008}) \bibinfo{pages}{1}.

\bibitem[{Armstrong et~al.(2004)}]{AGS2004}
\bibinfo{author}{T.~A. Armstrong}, et~al., \bibinfo{journal}{Phys. Rev. C}
  \bibinfo{volume}{70} (\bibinfo{year}{2004}) \bibinfo{pages}{024902}.

\end{thebibliography}
\end{document}